\documentclass[
 floatfix,
 aps,
 prl,
 amsmath,
 amssymb,
 superscriptaddress,
 longbibliography,
 reprint,%
]{revtex4-2}

\usepackage{graphicx} 
\usepackage{dcolumn} 
\usepackage{bm} 

\usepackage[utf8]{inputenc}
\usepackage[T1]{fontenc}
\usepackage{mathptmx}
\usepackage{mathtools}
\usepackage{relsize}
\usepackage{braket}
\usepackage{etoolbox}
\usepackage{hyperref}
\usepackage{natbib}
\usepackage{microtype}
\usepackage{booktabs}
\usepackage{hhline}
\usepackage{xcolor}
\usepackage{titlesec}
\usepackage{afterpage}  
\usepackage{mathrsfs}
\usepackage{siunitx}
\usepackage{flafter}

\titlespacing\section{0pt}{12pt plus 4pt minus 2pt}{4pt plus 2pt minus 2pt}
\titlespacing\subsection{0pt}{12pt plus 4pt minus 2pt}{4pt plus 2pt minus 2pt}
\begin{document}
\preprint{APS/123-QED}

\renewcommand{\topfraction}{.99}
\renewcommand{\floatpagefraction}{.99}

\title{Ionization and temperature measurements in warm dense copper using x-ray absorption spectroscopy}

\author{T. Cordova}
  \email[contact email:]{tcordova@ucsd.edu}
  \affiliation{Lawrence Livermore National Laboratory, Livermore, CA, USA}
  \affiliation{Mechanical and Aerospace Engineering, University of California San Diego, San Diego, CA}
\author{E. V. Marley}
  \affiliation{Lawrence Livermore National Laboratory, Livermore, CA, USA}
\author{D. A. Chin}
  \affiliation{Laboratory for Laser Energetics, University of Rochester, Rochester, NY, USA}
\author{R. A. London}
\author{H. A. Scott}
\author{M. K. G. Kruse}
\author{T. D{\"o}ppner}
  \affiliation{Lawrence Livermore National Laboratory, Livermore, CA, USA}
\author{F. N. Beg}
  \affiliation{Mechanical and Aerospace Engineering, University of California San Diego, San Diego, CA}
\author{F. Coppari}
\author{M. Millot}
\author{J. Emig}
  \affiliation{Lawrence Livermore National Laboratory, Livermore, CA, USA}
\author{S. B. Hansen}
  \affiliation{Sandia National Laboratories, Albuquerque, NM, USA}
\author{P. M. Nilson}
  \affiliation{Laboratory for Laser Energetics, University of Rochester, Rochester, NY, USA}
\author{P. Sterne}
  \affiliation{Lawrence Livermore National Laboratory, Livermore, CA, USA}
\author{M. J. MacDonald}
  \affiliation{Lawrence Livermore National Laboratory, Livermore, CA, USA}

\date{\today}

\begin{abstract}
  We detail experimental results inferring ionization and temperature for warm dense copper plasmas at several times solid density (\qtyrange[range-units=single]{15}{25}{\gram\per\cubic\centi\meter}15) and temperatures of \qtyrange[range-units=single]{10}{21}{eV}. Experiments performed at the OMEGA Laser Facility generate uniform warm dense matter conditions via symmetric shock compression of a buried copper layer. The plasma is probed with a laser-generated x-ray source to collect the K-shell x-ray absorption spectrum. Fitting bound-bound absorption contributions from constituent charge states of copper provides an estimated $\overline{Z}$ of approximately \numrange{4}{7} for these warm dense copper plasmas. We find that these partially ionized plasmas have K-edge shifts of \qtyrange[range-units=single]{12}{30}{eV} and bound-bound resonance 1s$\rightarrow$3p absorption shifts of \qtyrange[range-units=single]{4}{26}{eV} with respect to the cold K-edge. This study provides necessary experimental data to improve ionization and opacity models in the warm dense matter regime.
\end{abstract}
\maketitle

\section{\label{sec:Intro}Introduction}
  Warm dense matter (WDM) is a challenging and complex regime that exists between that of ideal plasmas and condensed matter, where thermal, Coulomb, and Fermi energies are all of comparable magnitudes. Thus, the typical approaches to model and measure properties in the limiting regimes of ideal plasmas and condensed matter cannot be extended to WDM. A more complete understanding of WDM is important because of its central role in the dynamics of many high-energy-density systems such as inertial confinement fusion \cite{glenzer_observations_2007,zylstra_burning_2022} and in the interiors of astrophysical bodies\cite{osterbrock_internal_1953, guillot_interiors_1999, lutgert_platform_2022}. In such systems, a hot central core exists at extreme densities and temperatures, surrounded by a cooler dense shell of WDM. The equation-of-state (EOS) and opacity of the WDM region can impact the hydrodynamics and energy transport through the system \cite{hu_first-principles_2014}. In the WDM regime, ionization plays a critical role in determining the EOS, as the ionization state strongly affects the internal energy of the system. Due to the complex nature of models in this regime, collecting high-quality experimental data is essential for improving the predictive capabilities of current models.

  Accurate measurements of WDM are difficult due to several experimental challenges. First, generating WDM in the laboratory often includes significant temporal and spatial gradients that complicate analysis of the experimental observables. Incorporating gradients in the analysis of experimental data, while possible, increases the uncertainties in the inferred plasma conditions\cite{macdonald_quantifying_2022,liedahl_x-ray_2025}, thus a design limiting gradients is key for high-quality measurements in this regime. Further, the high densities of WDM generally require x-rays to  spectroscopically probe the plasma; these plasmas are also not hot enough to produce appreciable self-emission for x-ray diagnostic purposes. These challenges require diagnostics that use an external x-ray source such as x-ray radiography, x-ray Thomson scattering, and x-ray absorption spectroscopy (XAS), which have been employed in experiments to measure WDM properties \cite{cho_electronic_2011,hansen_fluorescence_2018,macdonald_colliding_2023,bishel_ionization_2023}. In the case of these experimental methods, a large uniform volume is desirable for high-quality measurements, and in the case of x-ray radiography and x-ray Thomson scattering, a fit with a full physics model is required to infer plasma properties from the measured signals. A uniform WDM volume measured with XAS can provide a robust means to infer important properties of the plasma, such as temperature and ionization, without reliance on a full physics model to forward fit to the spectrum. Here, we present experimental results at the OMEGA Laser Facility\cite{boehly_initial_1997} where a uniform warm dense Cu plasma was generated and measured using XAS at WDM conditions.
  
  Our experiments use direct laser ablation to launch two counter-propagating shocks that heat and compress a buried layer of Cu to WDM conditions. The counter-propagating shocks generate a uniform volume of WDM for a short time after they have propagated completely through the buried layer and rebounded.  Similar counter-propagating shock techniques have been demonstrated on other experimental platforms, although under different conditions and scientific objectives\cite{fortmann_measurement_2012,fletcher_ultrabright_2015, bailie_k-edge_2023}. An x-ray backlighter is timed to probe the sample during this uniform phase, enabling x-ray absorption spectroscopy measurements of the WDM layer.

  At these high-density conditions, existing atomic and plasma models have difficulty reproducing the observed absorption spectra, motivating the development of an approach to infer temperature and ionization by characterizing spectral features near the K-edge. Additionally, the presence of bound-bound 1s$\rightarrow$3p resonance absorption is used to infer the charge state distribution of Cu. From simulation and spectral diagnostics, we infer the buried Cu foils reaches several times solid density (15 to 25 g/cm$^3$) and temperatures ranging from 10 to 21 eV, and we find an average ionization of 4--7 across the achieved conditions. This study provides necessary experimental data to improve the predictive capabilities of models in the WDM regime.

\section{\label{sec:Motivation}Modeling Ionization of Warm Dense Copper}
  Ionization is a fundamental process in any plasma system and is especially important to understand in the WDM regime because the dense plasma environment modifies the embedded atoms' bound states and consequently the ionization balance. Generally, the dense plasma lowers the effective ionization potential of the ions, due to increased screening from the surrounding high electron densities and lowering of the continuum due to close neighboring ions. This can impact the EOS through changes in ionization, and consequently other state variables such as temperature and pressure. Ionization also impacts the opacity, as the ionization states present in the plasma will have different absorption cross-sections with bound-bound and bound-free absorption at characteristic energies affecting radiation transport through the system. An accurate model of the EOS and opacity in this regime requires a detailed knowledge of the ionization processes as a function of density, pressure, and temperature. EOS and opacity calculations depend on the atomic model used to account for the dense plasma environment. These models in the WDM regime cannot include all the relevant atomic states and truncations are made to make the calculations tractable. Experimental data is necessary to test the models and assumptions used in creating an EOS or opacity table in WDM. X-ray absorption spectra measured in these experiments exhibit strong K-shell absorption features that encode key information about the constituent ions in the plasma as well as the effects of the high densities on the resulting spectra. These results can be compared against model approaches that estimate the density effects in the plasma and to test different theoretical approaches.
  
  For a solid density sample at zero temperature, the absorption around the K-edge depends on the relevant charge state's 1s binding energy, the chemical potential (sensitive to the electron degeneracy), and the ionization potential depression of the system. As the system is heated, excited states will open channels for bound-bound resonance absorption, notably the 1s$\rightarrow$3p transitions at several eV temperatures and 1s$\rightarrow$2p transitions at >100 eV temperatures for Cu. At warm temperatures (1--100 eV), the plasma's charge state distribution is narrow, and each constituent ion will, when probed with external x-rays, contribute to the K-edge or resonance absorption at an identifiable energy. Increases in density can lead to additional ionization, and the ionization-dependent 1s binding energy will further shift the absorption features. Additionally, broadening effects of the observed absorption and emission lines from the surrounding plasma will affect on the inferred width of each contribution to the charge state distribution and thus should be included, but this will not affect our ability to infer the average ionization $\overline{Z}$.
    
  Calculating the 1s$\rightarrow$np transition energies typically begins with an isolated atom model, and the code used for this analysis was Flexible Atomic Code (FAC)\cite{gu_flexible_2008}, and a density-dependent model is added on top to account for the plasma environment. Simple models of the density-dependent plasma polarization shift (PPS)\cite{nguyen_atomic_1986} predict that at electron densities of $n_e\approx\qtyrange[range-exponents=combine,range-units=single]{1e24}{2e24}{\per\cubic\centi\meter} $, the 1s$\rightarrow$3p lines shift by a few to 10 eV, and 1s$\rightarrow$2p will shift a few eV. With consideration of these shifts, we make a direct inference of the constituent charge state distribution (CSD) of the Cu plasma. 
  
  Comparing experimental data with different models requires well-constrained plasma conditions in phase space, including temperature, density, and the (CSD), to accurately capture density-dependent effects in WDM. Measuring the K-edge with XAS can offer a powerful and robust means to infer temperature conditions within a WDM plasma independent of the material choice. The absorption spectrum at the K-edge is strongly characterized by the photoelectric coupling of a 1s electron to a vacant near or in the continuum. The photoabsorption cross section can be written as
   \begin{equation}
    \label{eqn:pixs}
    \sigma(h\nu) = ah\nu\left\|\bra{\psi_{1s}}\|R\|\ket{\psi_{f}}\right\|^{2}[1 - f(E)]
   \end{equation}

  where R is the electric dipole moment operator acting on the initial 1s state, $\bra{\psi_{1s}}$, and selecting a p state among the final states, $\ket{\psi_{f}}$. Under degenerate conditions, Fermi-Dirac statistics govern the degenerate electron population, and the probability of a vacancy in the continuum near the K-edge, $[1 - f(E)]$, will be the dominant contributor to the energy profile of the K-edge. Here $f(E)$ is the electron occupancy, which follows a Fermi-Dirac distribution:

    \begin{equation}
      \label{egn:FDD}
      f(E) = \left[1 + \exp\left[\frac{E - \mu(T, \epsilon_f)}{k_B T}\right]\right]^{-1}
    \end{equation}

  where E is the energy, $\mu$ is the chemical potential, T is the temperature, and $k_B$ is the Boltzmann constant. The Fermi energy is $\epsilon_{F} = \hbar / 2m_e [3\pi^{2}n_e]^{2/3}$, where $n_e$ is the electron density, and $m_e$ is electron mass. 
  
  Inference of temperature from the edge requires the material to be sufficiently degenerate, such that the Fermi energy exceeds the electron temperature. Under this condition, the temperature can be directly read independently of the energy position of the edge. Potential contributions to the K-edge shape from the initial state are insignificant, as $\bra{\psi_{1s}}$ is largely unaffected in this WDM regime. Meanwhile, the final state $\ket{\psi_{f}}$ implicitly contains a dependence the electronic density of states (DOS). Where here, the DOS is calculated from the square modulus of the final state integrated over space. So long as the profile of the DOS remains slowly varying over the K-edge region, the inference of temperature at the slope is reliable. Previous work by Dorchies et al. \cite{dorchies_x-ray_2015} has explored the sensitivity of the K-edge to the temperature of Al and has shown that for T > $\epsilon_F / 4$ other effects such as a varying DOS or thermal broadening will affect the K-edge shape and the reliability for direct temperature readings. The K-edge will progressively underreport the temperature, signifying our inference as a lower bound of temperatures in conditions closer where $k_B T$ and $\epsilon_F$ are of a comparable value. A further effect that could be expected due to the density of states is the presence of other absorption features, 1s$\rightarrow$3d, though we can expect these non-dipole transitions to be far less probable than the 1s$\rightarrow$np transitions.

  \begin{figure}
    \centering
    \includegraphics[width=\linewidth]{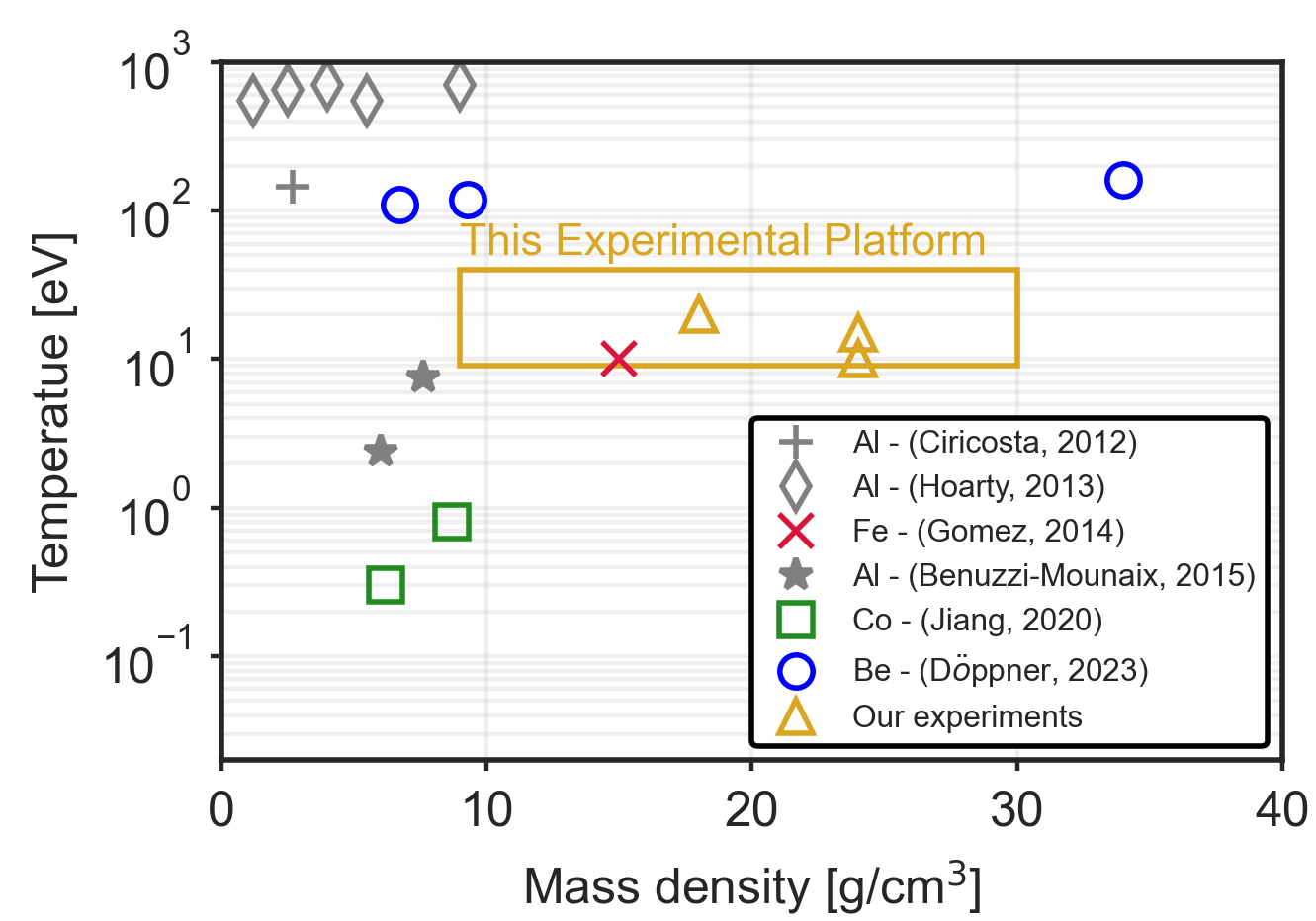}
    \caption{Comparison of this experiment's capabilities to some previous experiments that focus on measuring ionization and density effects in WDM. The experimental configuration in this study is capable of achieving $\rho\approx$ 10--30 g/cm$^{3}$ at T $\approx$ 10--30 eV for Cu.}
    \label{fig:phasespace}
  \end{figure}
  
  Consideration of how the dense plasma environment lowers an ion's effective ionization potential, resulting in so-called ionization potential depression (IPD), is required to infer ionization information from the K-edge. The IPD, electron degeneracy, and ionization all cause changes to the observed energy and profile of the K-edge. Two of the most commonly used IPD models are the Stewart-Pyatt (SP) model \cite{stewart_lowering_1966} and the Ecker-Kroll (EK) model\cite{ecker_lowering_1963}. The SP model interpolates from the Debye-Hückel model in the low-density limit to the ion-sphere model in the high-density limit. The EK model is derived from microfield potential contributions near the ion and Coulombic interaction in the far field. The notable distinction lies in incorporating the total particle density, $n_i$ + $n_e$, in the EK model leading to typically stronger IPD than the SP model. These ad-hoc modifications have been shown to describe plasmas in specific regimes, but there have been varying results favoring each IPD model, or entirely different approaches (i.e. density-functional theory models), in different regimes as described below.

  There have been efforts in recent decades to experimentally measure and characterize warm dense metals. Figure \ref{fig:phasespace} shows some key experiments in this regime of WDM that investigate dense plasma effects. The previous experiments can be categorized based on their experimental method: XFEL experiments shown in Figure \ref{fig:phasespace} from Ciricosta et al. \cite{ciricosta_direct_2012,ciricosta_measurements_2016} isochorically heats targets to 100 to 200 eV at solid densities; Short pulse laser systems are also able to achieve extreme temperatures ($\approx$500 eV) at solid density as demonstrated by Hoarty et al. \cite{hoarty_observations_2013}; Pulsed power experiments by Gomez and Hansen et al. \cite{hansen_fluorescence_2018} measured Fe in MagLIF liners\cite{gomez_experimental_2014} at Z Machine achieving 15 g/cm$^3$ and 10 eV; long pulse laser systems can shock heat targets, experiments by Benuzzi-Mounaix et al. \cite{benuzzi-mounaix_electronic_2011} show capabilities of 3x solid density and 8 eV temperatures. Conversely, ramp compression has been employed by Jiang et al. to compress Co to 1.5--2.1x solid density at lower temperatures < 1 eV; Finally, experiments by D{\"o}pper et al. \cite{doppner_observing_2023} measured ionization of Be at extreme conditions ($\rho\approx$ 7--55 g/cm$^3$, T$\approx$110--160 eV) with capsule implosions on the NIF\cite{spaeth_description_2016}. Importantly, among these experiments shown, no agreement exists on the IPD model that best describes the ionization in this regime of warm dense plasmas. The experiments presented in this study span an area of the phase space denser than shown with XFEL or short pulse laser experiments, and moderate temperatures that have little experimental data with a mid-Z metal. Our results provide additional data points that are, importantly, designed to be uniform plasmas and yield a valuable constraint to any model describing ionization in warm dense plasmas.

  \begin{figure*}
      \centering
      \includegraphics[width=\linewidth]{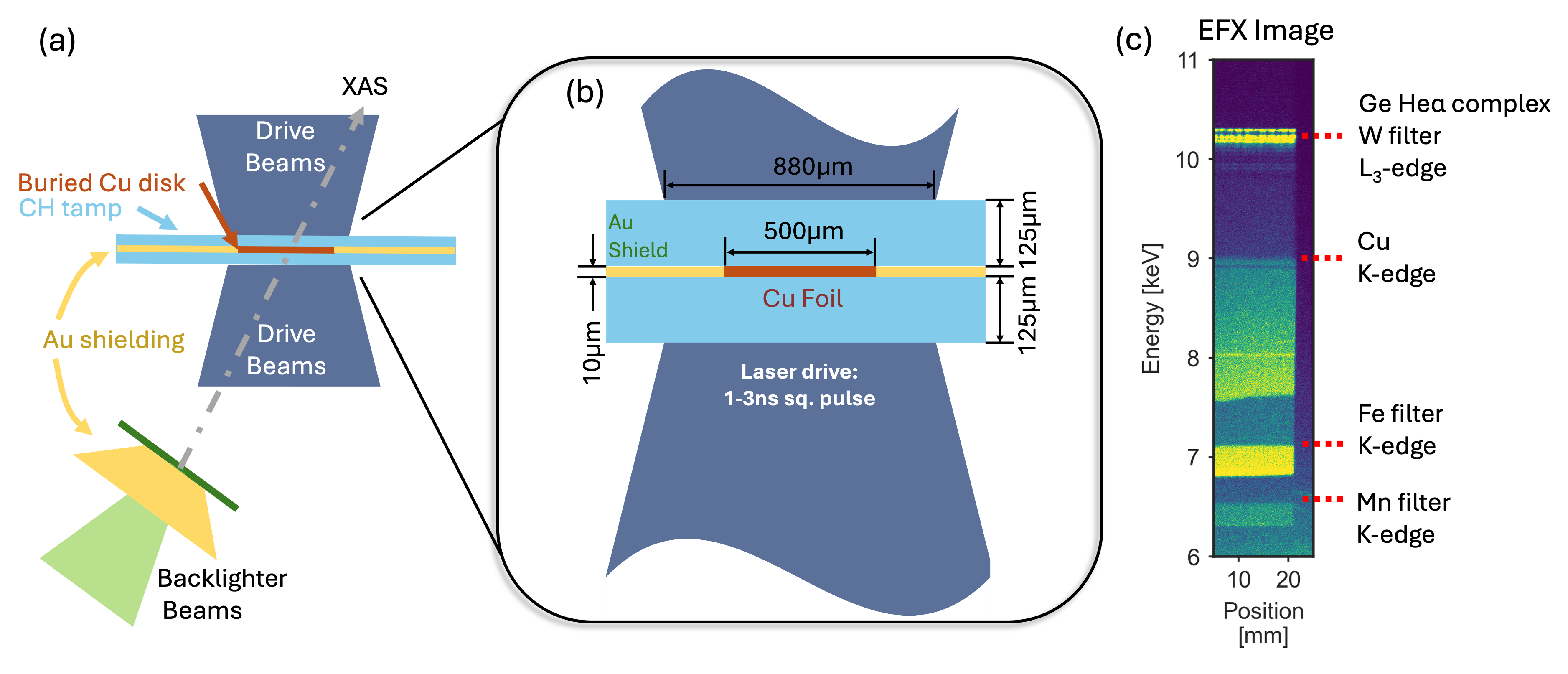}
      \caption{(a)This experimental configuration consists of the main target of CH, 125 $\mu$m, and buried Cu foil surrounded by Au washer, 10 $\mu$m thick; and a Ge backlighter. Shown are the representative drive and backlighter laser beams and the spectrometers (XAS, XRF) lines-of-sight. (b) Detailed sizes of the target layers. The laser drive is configured with a 3-1ns sq. pulse with ~3400-5900 J on target per side at a spot size of 880$\mu$m. (c) Example XAS image from experiment.}
      \label{fig:omegaplatform}
  \end{figure*}
  
  \begin{figure*}[!t]
    \centering
    \includegraphics[width=0.95\textwidth]{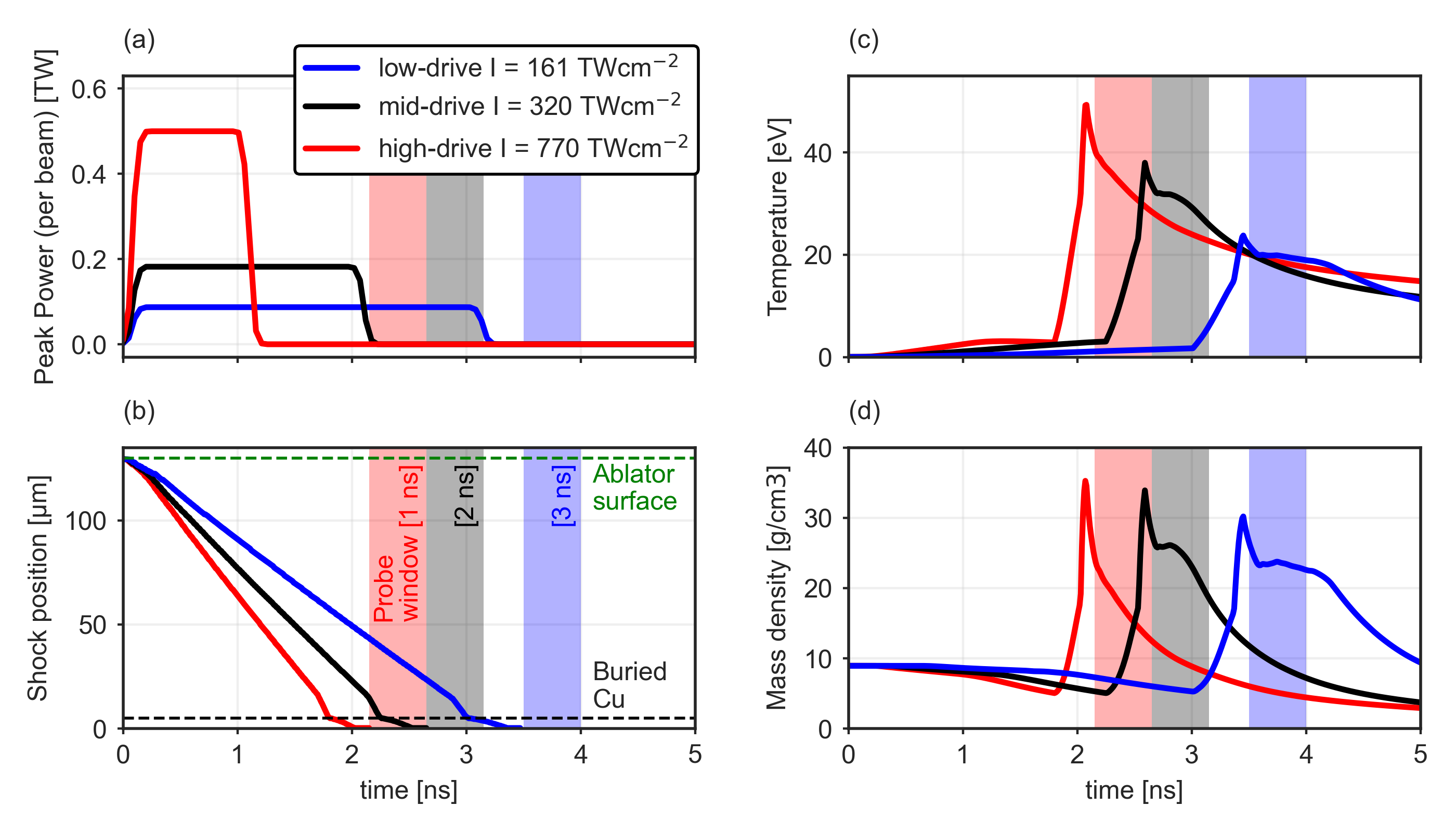}
    \caption{(a) Laser intensities for each drive configuration. (b) Shock trajectories as modeled by 1D HYDRA simulations for each drive. (c) Electron temperature within the Cu layer over time and (d) shows mass over time for each. The shaded regions represent the optimal backlighting time window, for each configuration, for probing a uniform warm dense volume of copper.}
    \label{fig:hydraplots}
  \end{figure*}

\section{\label{sec:exp_setup} Target Design and Experimental Configuration}
  Experiments were conducted at the OMEGA Laser Facility, which is a 60 beam laser able to focus up to 30 kJ of 351 nm light over 1--3 ns on target. The experimental configuration consists of a planar target and a separate backlighter positioned 3 mm away.  At the center of the planar buried layer target is a 10 $\mu$m thick Cu disc (8.96 g/cm$^3$ solid density) with a 500 $\mu$m diameter. This is surrounded by an Au washer cut to match the Cu and acts as an x-ray block for the x-ray diagnostics. The Cu and Au lie between two symmetric plastic (CH) ablators, each with a thickness of 125 $\mu$m. A diagram of the target package is shown in Figure \ref{fig:omegaplatform}(a). The backlighter is a 6 $\mu$m thick Ge, coated onto a 100 $\mu$m graphite foil. Additionally, the backlighter has shielding to prevent emission from the expanding laser plume from being in the line of sight of any diagnostic.
  The laser drive uses 14 beams on each side of the primary target, producing 3.4 to 5.4 kJ total laser energy on target per side, configured with a standard square pulse of either 1, 2, or 3 ns duration to achieve a variation of shock strengths that compress and heat the Cu sample. The achieved laser intensity is 161--770 TW/cm$^{2}$ over the three pulse length configurations. The laser spot is smoothed by with distributed phase plates (DPPs) for each beam and smoothing by spectral dispersion (SSD) to reduce hot-spots in the drive and improve uniformity. The laser spot is chosen to overfill the target such that the buried layer sees a planar shock across its entire diameter, beginning from a uniform beam of 880 $\mu$m diameter shown in Figure \ref{fig:omegaplatform}(b). The Ge backlighter is separately timed and irradiated with six additional beams, which deposits approximately 1.2 kJ total laser energy in a 500 ps square pulse at a spot diameter of 140 $\mu$m. 

  \begin{figure*}[t]
    \includegraphics[width=\linewidth]{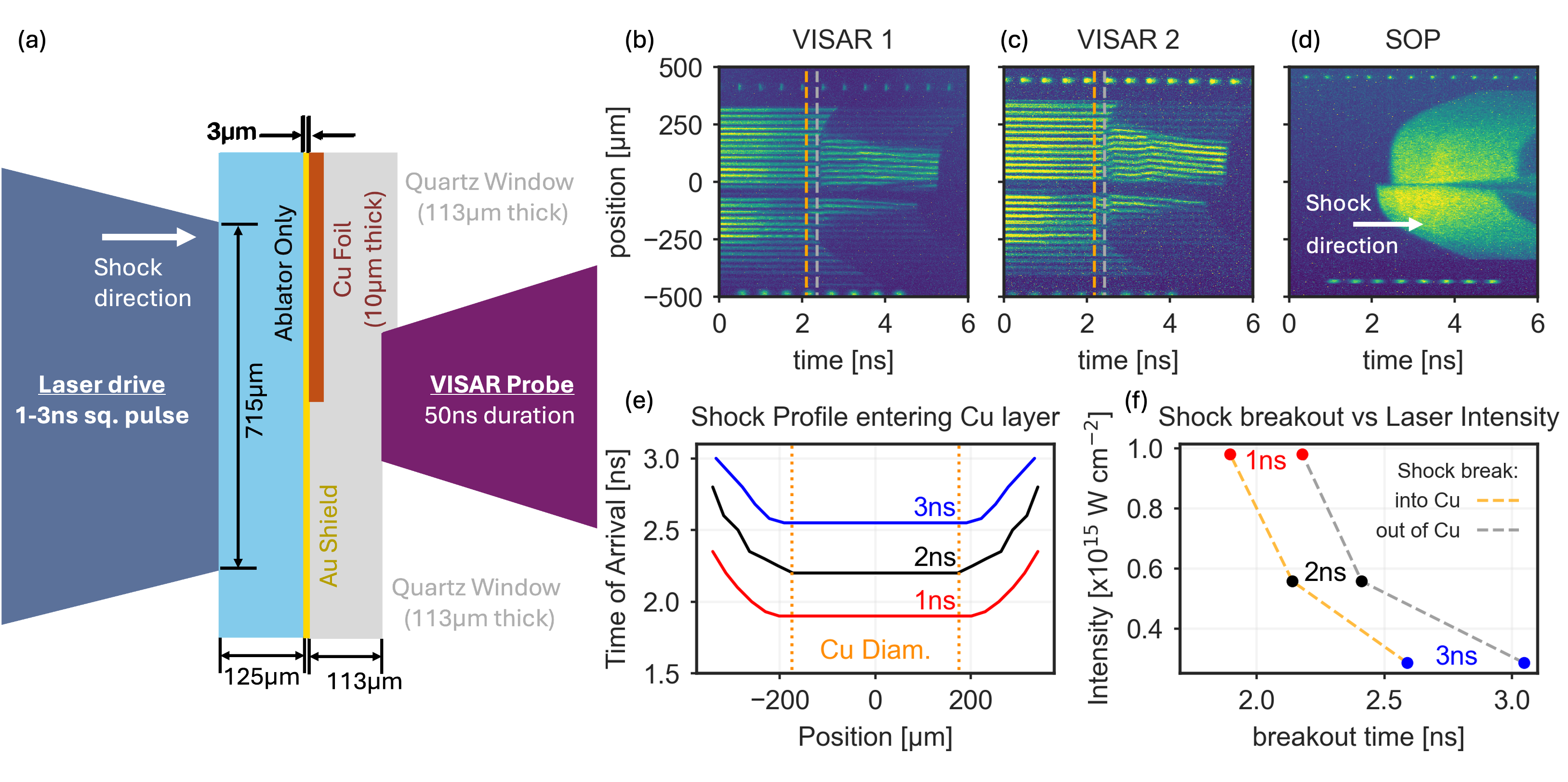}
    \caption{(a) Target schematic of experiments for measuring shock timing with the VISAR. VISAR measurements for the mid-drive (2 ns pulse length) configuration, two VISAR legs (b) and (c) and SOP (d) measure the shock breakout of the Cu layer (upper) ablator (lower); the breakouts are denoted by the dashed lines for the ablator (orange) and Cu (gray). (e) The shock profiles for each drive configuration as inferred from 2D VISAR data as it enters the Cu layer, as compared to the Cu foil diameter (dashed orange). (f) the relationship of shock timing into, and out of, the Cu layer vs. the driven laser intensity, this is used to aid in timing two-sided experiments for probing uniform conditions.}
    \label{fig:visar_exp}
  \end{figure*}

  The 500 ps backlighter duration sets the timing window of our time-integrated diagnostics and is timed relative to the main laser drive to probe when the plasma achieves uniform conditions just after the symmetric shocks propagate through the buried sample. Figure \ref{fig:hydraplots}(a) depicts the three pulse length configurations used in this study. Dedicated experiments were completed with one-sided versions of the targets with an added quartz window to measure the shock breakout with VISAR\cite{celliers_imaging_2023} at the same laser intensities. Analysis of the experiments dedicated to measuring the shock timing with VISAR is further detailed below, in Section \ref{subsec:visar}. We used the multi-physics radiation-hydrodynamics code HYDRA\cite{marinak_comparison_1998} to simulate both the one- and two-sided experiments. We determined constant laser power multipliers for the one-sided experiments that match the shock timing measured with VISAR, yielding the shock trajectories shown in Figure \ref{fig:hydraplots}(b). These multipliers were then used  to achieve the same laser intensities (per side) on the two-sided HYDRA models to ensure proper backlighter timing and to predict the temperatures and densities achieved in the Cu layer. The HYDRA predictions for the average temperature and density over the Cu layer are shown in Figure \ref{fig:hydraplots}(c) and (d), respectively. The backlighter timing windows over which the XAS measurements are made for each pulse are shown as shaded regions in Figure \ref{fig:hydraplots}. An accurately timed backlighter ensures a more uniform plasma for the measured XAS and for the inference of temperature from the Cu K-edge. The longer pulse durations provide more uniformity over the timing window, whereas the high-drive (1 ns pulse) is the least uniform over the measurement duration.

  Our primary diagnostic was XAS, obtained using the EFX spectrometer\cite{chin_high-resolution_2023}, which measures the K-edge x-ray absorption spectrum over the backlighter timing window. The EFX diagnostic was configured with a flat Si (111) Bragg crystal to cover an energy range from 6.3 to 11.4 keV. The XAS signal was recorded on image plate using a filter pack consisting of Mn, Fe, and W, which provides fiducial K- and L-edge features for fitting the spectral dispersion on each shot. These filter edges are used in combination with a ray-trace model implemented in the XICSRT code\cite{pablant_design_2021} to constrain the energy dispersion relation to within $\pm$3.5 eV.

  For the backlighter source size used in these experiments (140 $\mu$m diameter), we achieve a spectral resolution of approximately 10 eV at 9 keV ($E / dE = $900), with source-size broadening providing the dominant limitation. Appendix \ref{appendix:efx_res} details the theoretical framework used to estimate the spectral resolution for this experimental configuration. In practice, changes in the K-edge slope corresponding to temperature differences greater than $\sim$3 eV are resolvable, and the minimum temperature distinguishable from the cold case is approximately 2 eV. A representative EFX image from these experiments is shown in Figure \ref{fig:omegaplatform}(c).

\section{\label{sec:exp_results}Experimental Data and Analysis}

   \subsection{\label{subsec:visar}Experimental VISAR measurements}
    Dedicated experiments were conducted on a one-sided version of the buried layer targets to measure the shock break in and out of the Cu layer using the line-imaging VISAR\cite{barker_laser_1972} diagnostic to improve timing accuracy of the backlighter and sizing of the target diameter.

    Figure \ref{fig:visar_exp}(a) shows a basic schematic of the one sided-targets, and the relevant size of the laser in this configuration. The targets are of a similar construction to the two-sided targets, with a 125µm thick CH ablator  and 10µm Cu foil beneath. Some changes specific to improve the VISAR measurement include a 3µm thick Au foil between the ablator and Cu, this is to prevent preheating and photoionization to ensure the VISAR quartz window remains transparent throughout the measurement. The Cu is also placed over only one half of the target to get a direct measure of both the breakout times from the ablator with Au heat shield and the Cu for each shot.  

    Each experiment has two VISAR channels with different sensitivities; one leg was set with 33.66 $\mu$m/ns/fringe, and the second with 13.538 $\mu$m/ns/fringe, to resolve fringe jump ambiguities to determine the shock velocity time history as a function of position across the target measured within the quartz window. Additionally, a streaked optical pyrometer (SOP) is employed on the same line-of-sight, which gives another indication of shock breakout in the quartz. Figure \ref{fig:visar_exp}(b) shows an example dataset from the 2ns pulse configuration. The shock breakout into the quartz window for both ablator with Au heat shield only and with Cu can be seen at 2.15 ns and 2.40 ns, respectively, by the shift in the fringes in \ref{fig:visar_exp}(b) and \ref{fig:visar_exp}(c). The line-imaging capability of the VISAR at OMEGA also allows for a determination of the shock planarity across the buried layer target. The inferred transverse shock profile for all three pulse configurations is shown in Figure \ref{fig:visar_exp}(c); this is extracted from the discontinuity of the fringes transverse to the shock direction. This gives a quantitative estimate of how large the planar region of the shock is for the laser configuration. The Cu layer is then chosen to be well within this planar area to best maintain uniformity for the XAS measurements. Figure \ref{fig:visar_exp}(e) shows that for these VISAR experiments that a target diameter of 350 µm or less will see little variation in the planarity of the shock. The two-sided experimental iterations used an increased spot size to allow for a 500 $\mu$m diameter foil while maintaining the same planarity of the propagating shock and the Cu region. 

    The relationship between the laser intensity and the measured shock breakout is shown in Figure \ref{fig:visar_exp}(f). The timing information is then used to fit the 1D HYDRA simulations with the scalar multiplier to the intensity. We found each drive configuration required its own multiplier to fit the VISAR timing, and these multipliers were used to predict the timing window for the two-sided symmetric drive experiments detailed in this study. The multipliers needed to scale HYDRA to match the measured shock timing were 0.75, 0.56, and 0.41 for the low-, mid-, and high-drive configurations, respectively, and these calibrated simulations directly informed the timing of the backlighter in the primary XAS experiments.

  \subsection{\label{sec:exp_infer}Extracting Inferred Temperature and Ionization from the Cu K-edge}
\begin{figure}
      \centering
      \includegraphics[width=1.0\linewidth]{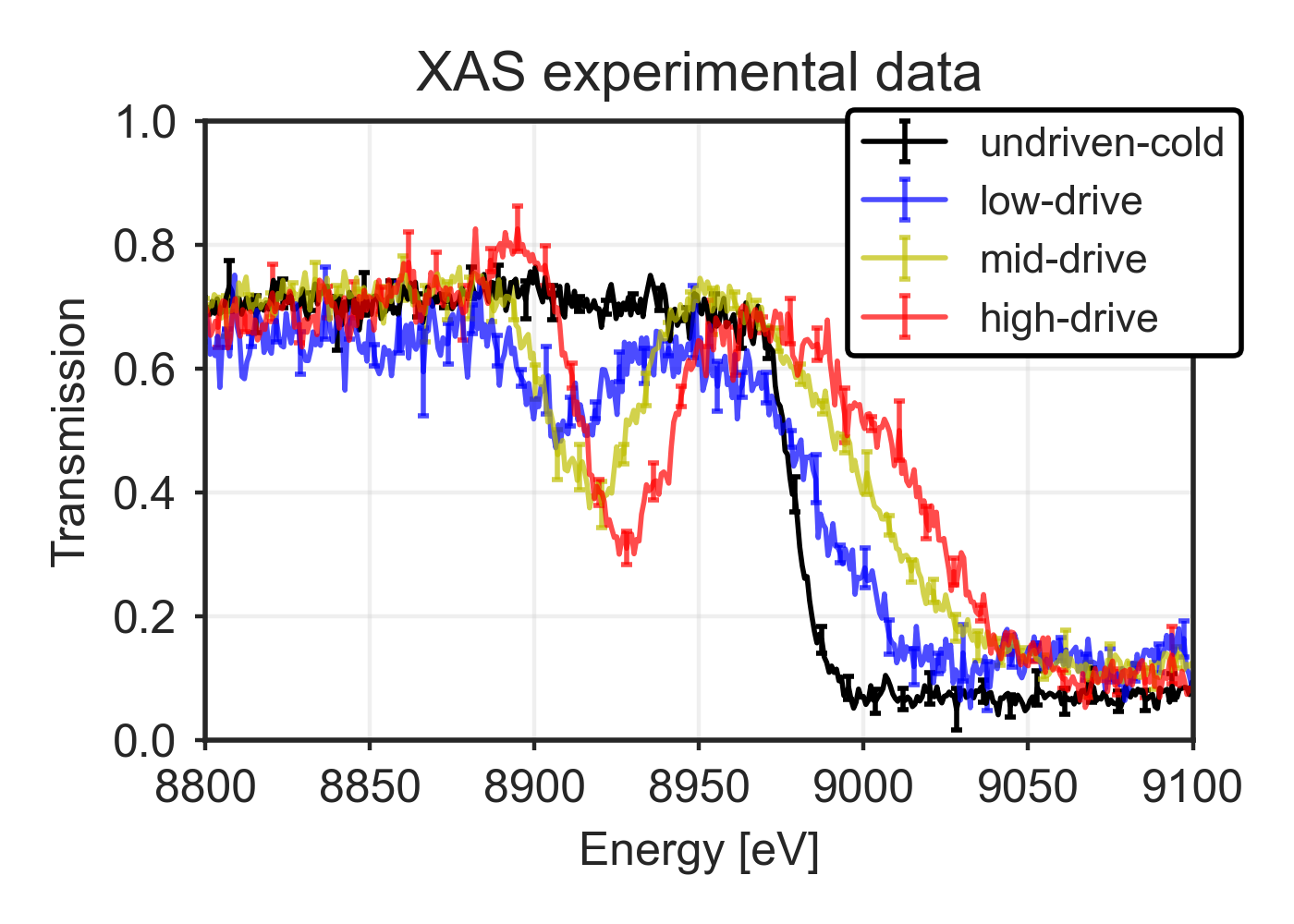}
      \caption{XAS data collected at the OMEGA laser facility. Experiments varied laser drive intensity to heat and compress the Cu buried layers. In black is the ambient case and low- (blue), mid- (orange) and high- (red) drives are of increasing intensities.}
      \label{fig:allshots}
    \end{figure}

    The measured Cu XAS from these four experimental configurations are shown in Figure \ref{fig:allshots} denoted throughout as undriven, low-drive, mid-drive, and high-drive configurations corresponding to no laser drive, 3ns drive, 2ns drive, and 1ns drive, respectively. These spectra qualitatively illustrate how the plasma conditions change, and therefore the K-edge, with drive intensity. Sensitivity to temperature in the K-edge and ionization in the resonance absorption show changes in the Cu plasma conditions as the drive is changed.  We also note the presence of K$\beta$ fluorescence pumped by the laser driven plasma, most notably in the high-drive case. Because the source (the ablation plasma) pumps the fluorescence from a different location and early in time, it is not expected to significantly effect the XAS measurement, but it can contribute to the K-edge region shape and that must be considered in the analysis.

\begin{figure}
      \centering
      \includegraphics[width=1.0\linewidth]{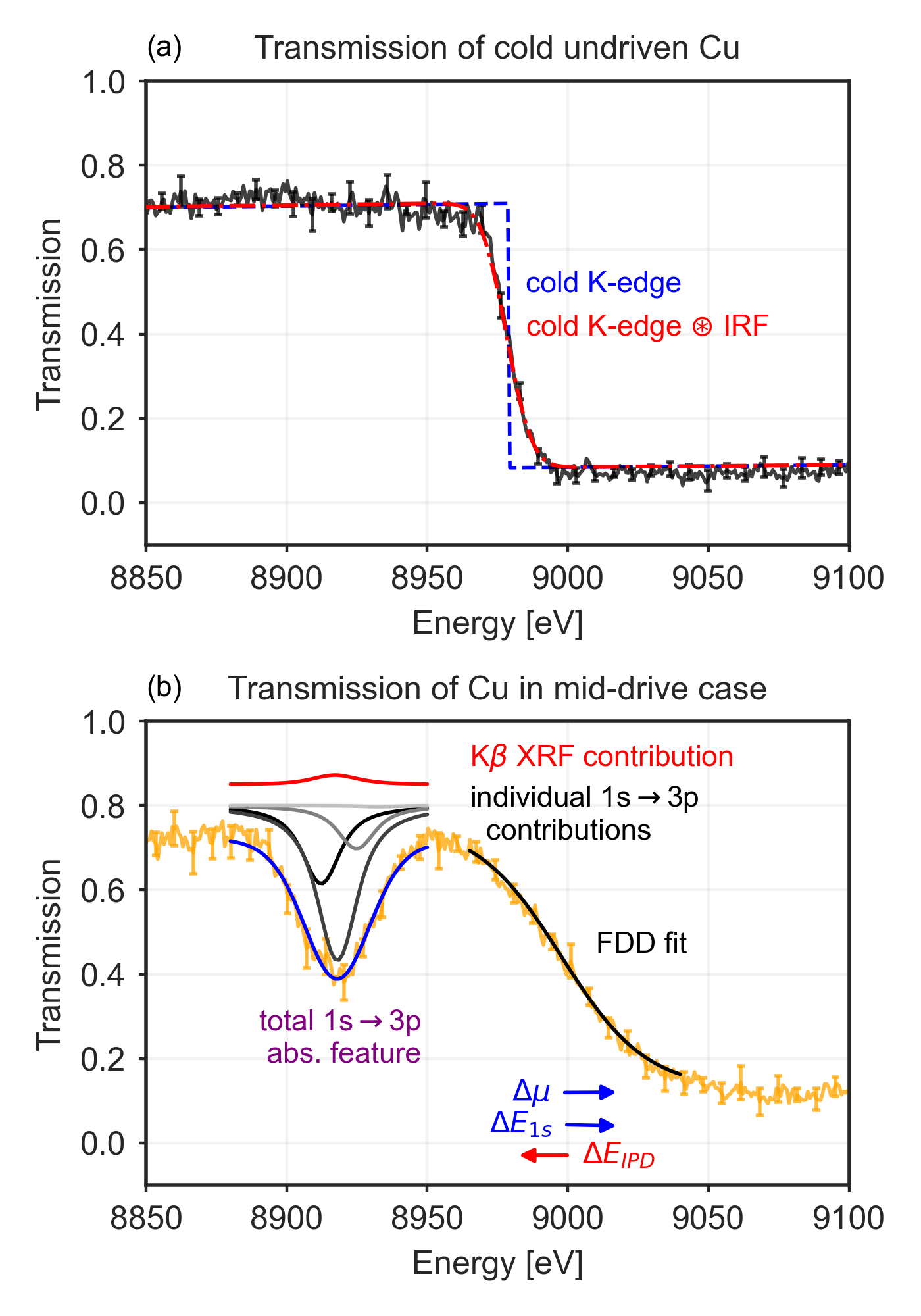}
      \caption{(a) Experimental spectra of cold Cu compared to tabulated cold transmission data. (b) Mid-drive experimental spectra overlaid with respective fits for the K-edge, bound-bound resonance absorption, and K$\beta$ emission. Contributions from each individual fitted charge state is shown in grayscale and the combined in blue. Together, this fit method extracts an inferred temperature and charge state distribution of the warm dense Cu. Factors that can affect the K-edge position are the 1s binding energy ($\Delta E_{1s}$), the chemical potential ($\Delta\mu$), and IPD ($\Delta E_{IPD}$), the direction at these conditions they can shift the edge are denoted by the arrows (not to scale).}
      \label{fig:ambanddrive}
    \end{figure}

    Accurate analysis of the K-edge slope and broadening of the 1s-3p absorption features require an accurate instrument response function (IRF) to account for the spectral resolution of the measurement. The IRF was determined from the undriven configuration that measured the cold Cu K-edge. The total instrument response here includes the inherent instrument resolution as well as source broadening from the backlighter size, in the case of absorption, and from the Cu target size, in the case of K$\alpha$ and K$\beta$ emission. The backlighter source function can also be determined from this measured data, since the cold Cu density and thickness is known, which are important for characterizing and analyzing the driven shots. Figure \ref{fig:ambanddrive}(a) shows the undriven measurement compared to a tabulated cold edge \cite{schoonjans_xraylib_2011}.

    \begin{figure*}
      \centering
      \includegraphics[width=1.0\linewidth]{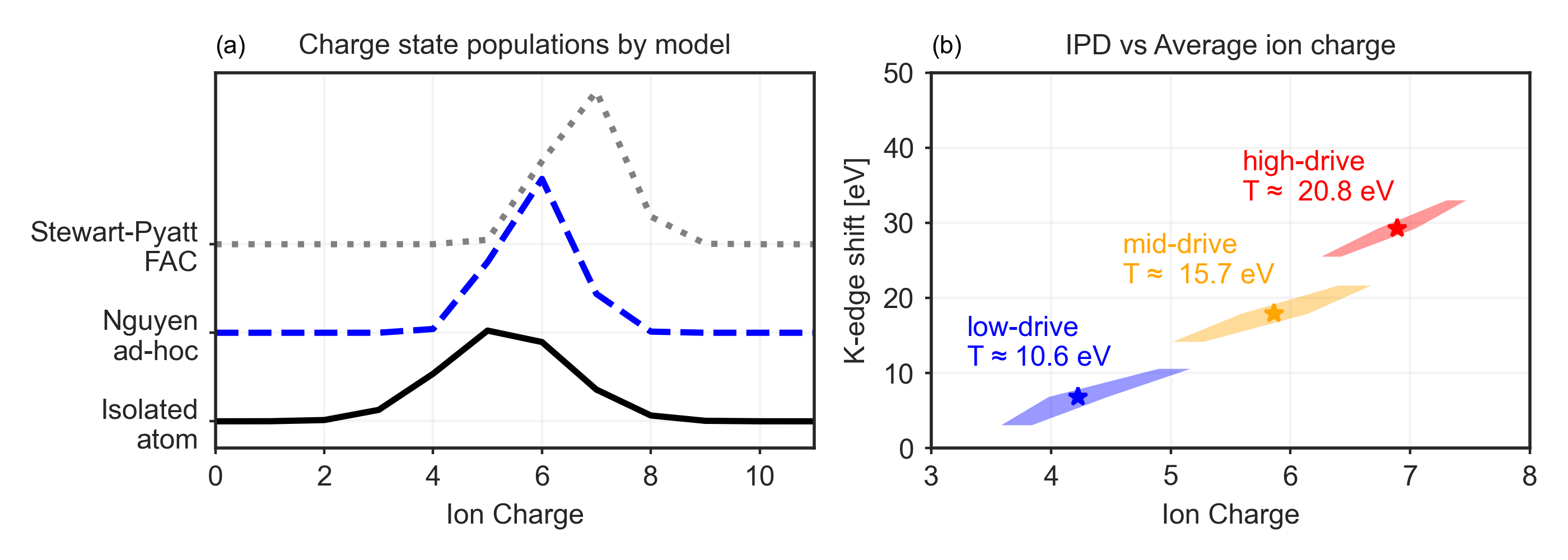}
      \caption{(a) Inferred charge state distribution with three models, each prescribe a different level of PPS in the warm dense Cu leading differing ionization in this mid-drive case. (b) Comparison of the K-edge shift from cold for each drive configurations at the inferred $\overline{Z}$ (Nguyen). See Table \ref{tab:summary} for detailed values.}
      \label{fig:csd_shifts}
    \end{figure*}
   
    Before the temperature and ionization can be inferred from the K-edge region, the signal contributions from the background laser drive and backlighter are removed. The wide energy range of EFX allows for the measured signal away from spectral features to be used to fit a combined background and source function consistent with our undriven measurements. The measured absorption around the K-edge is modeled by fitting to two major components: a Fermi-Dirac distribution that can capture the temperature sensitivity at the K-edge, and a set of 1s$\rightarrow$3p absorption lines from each contributing charge state to capture the absorption feature below the K-edge energy. Our fitting model utilizes a Markov Chain Monte Carlo (MCMC)\cite{foreman-mackey_emcee_2013} sampler to determine the value of the free parameters of interest to infer temperature and ionization, along with statistical uncertainties and correlations.
   
    Inferring the temperature from the Fermi-Dirac component requires four total parameters, three to position it at the K-edge (an amplitude $A_1$, a centering value $\mu_1$, and a y-offset $y_1$) and finally, a temperature $k_{B}T$ to fit the K-edge slope. The absolute position of the K-edge requires knowledge of each the 1s binding energy and the density of states for each ion along with the chemical potential of the system, and the IPD, and therefore requires a full physics model. However, because of the edge's dependence on the density of states in addition to the vacancies given by the Fermi-Dirac distribution, at higher temperatures this simple method obtains a lower-limit on the temperature, which we report here. In contrast, the HYDRA simulations tend to estimate much higher achieved temperatures; which could be a result of several mechanisms including the specific ionization model in the code.

    We fit the charge state distribution using four free parameters by assuming a mean ionization $\overline{Z}$ from which a super Gaussian can approximate the weighted contributions from the charge states surrounding $\overline{Z}$. The MCMC model can then adjust all 30 charge state weights by adjusting the Gaussian variables (amplitude $A_2$, variance $\sigma_{2}^{2}$, centering or $\overline{Z}$, and the Gaussian exponent $P$). To model the 1s$\rightarrow$3p absorption feature, we used Voigt profiles for each charge state, weighted by the relative population of each state as determined by the Gaussian weights. The Voigt width is calculated with both Stark and natural broadening, which will depend on the density and $\overline{Z}$; denoted as $\Delta_{br}$ below. The Stark broadening contribution is calculated using a model derived from Griem\cite{griem_plasma_1964} and natural widths have been estimated from Cretin\cite{scott_cretinradiative_2001} calculations but were found to be negligible in comparison to the IRF and Stark broadening of the absorption feature; the combined effects resulted in the fitted shaped seen in Figure \ref{fig:ambanddrive}.
   
    Fitting the Voigt functions to the measured signal requires a calculated spectral energy where each will contribute to the total absorption, for this we use FAC. Next, the choice of imposing a plasma polarization shift on the charge states is required, and for this, there are two options explored in this study. The first is a widely used ad-hoc model formulated by Nguyen et al. \cite{nguyen_atomic_1986}, from which PPS can be calculated at a density and temperature for the 1s$\rightarrow$3p transition. This is denoted by $\Delta_{PPS}$ above and is used to shift the isolated atom transition energies. We compare this to a more fundamental approach that adds plasma screening directly in the FAC transition calculations, removing the need for an additional PPS modifier.

    To calculate terms in the model such as the PPS or broadening that required an electron density, we used the density in combination with the found $\overline{Z}$ and this is varied within the model as a fitted solution of $\overline{Z}$ is converged to. There is no direct measure of density for this experimental data, however, our HYDRA simulations estimate the achieved densities of each drive, albeit with large uncertainties. The laser energy variation between our VISAR experiments to the four presented here (low--high drives and undriven) was also used to estimate the change in the simulated density values found with HYDRA. The simulations used for estimating the two-sided experiments utilize the found scalings from the VISAR but with the appropriate layers and the symmetric boundary for the counter propagating shock configuration. For these experiments, the laser varied < 10\% from the expected intensities and this variation in combination with potential timing uncertainty in the backlighter, $\pm$20 ps, is used to quantify the expected uncertainty in the simulated densities. HYDRA's underlying EOS tables were investigated, and we found utilizing either the LEOS or Sesame EOS tables\cite{mchardy_introduction_2018} does not incur a notable difference in the final densities found. To assess model sensitivity, we varied additional multiplicative parameters in HYDRA, including inverse bremsstrahlung (laser absorption), electron conductivity, the electron flux limiter, electron--ion coupling, radiative opacity, and artificial viscosity. Even for large deviations from nominal values (0.5--2x nominal), the resulting densities changed by less than 1\% for some parameters and by no more than $\pm$8\% for the most sensitive (inverse bremsstrahlung and electron conductivity), this has been included in the uncertainty of the densities. The densities found via HYDRA simulations are 24$\pm$3, 24$\pm$3, and 18$\pm$4 g/cm$^3$ for the low, mid, and high drives, respectively. Finally, there is a contribution from K$\beta$ fluorescence emission, which also depends on the charge state distribution; this can include emission pumped by the laser-driven plasma, as observed most clearly in the high-drive experiment. The k$\beta$ emission amplitude is scaled from measured from the K$\alpha$ emission on EFX using $\alpha$/$\beta$ ratios previously verified experimentally\cite{akkus_k-betak-alpha_2017}. The full fitting model then takes the form:

  \begin{equation}\label{eqn:fitmodel2}
   \begin{split}
      F_{tot} &= A_1\left[1 + \exp\left[\frac{E - \mu_1}{k_B T}\right]\right]^{-1} + y_1\\
      &+ \sum_{z=0}^{29} A_2 \exp\left[-\left(\frac{z - \overline{Z}}{\sigma_2}\right)^P\right] F_{Voigt}(E, z, \Delta_{br}, \Delta_{PPS}) \\ 
      &+ F_{K\beta}(E, \overline{Z})\\
   \end{split}
  \end{equation}
   
  Additionally, the curves are convolved with the IRF when fitting the experimental data. The total fitted model contains the sum of contributions from each charge state and temperature influence on the K-edge through the Fermi-Dirac distribution component. Figure \ref{fig:ambanddrive}(b) shows the final resulting fit split into its functional components for the mid-drive, the inferred temperature is $\approx$ 16 eV and a mean charge of $\approx$ 6--7, dependent on the model choice in calculating the 1s$\rightarrow$3p transitions.

  Together we have a means to infer the temperature and ionization of the buried Cu plasma without the need for a forward fitting physics model. Application of this fitting method yields inferred temperatures for the low, mid, and high drives of 10.6$\pm$0.4, 15.7$\pm$0.6, and 20.8$\pm$0.7 eV, respectively. Figure \ref{fig:csd_shifts}(a) also shows the variation in the inferred charge state distribution by model for the mid-drive. From the resonance absorption component, we find an average ionization $\overline{Z}$ of 4.2$\pm$0.8, 5.9$\pm$0.8, and 6.9$\pm$0.8 with ad-hoc PPS for the low, mid, and high drives, respectively. In comparison, FAC, with an applied Stewart-Pyatt potential as the plasma screening model, yields similar results shown in Figure \ref{fig:csd_shifts}(a) which is compared with the FAC isolated atom calculation in black. FAC with the Stewart-Pyatt potential model estimates the strongest IPD effect on the resultant CSD for the mid- and high-drives and slightly weaker shifts in the low-drive. The found $\overline{Z}$ for this PPS model is 4.0$\pm$0.8, 6.9$\pm$0.8, and 8.6$\pm$0.8 for the low-, mid-, and high-drives, respectively. Together with our density estimate, we have an inferred set of $\rho$, T, and $\overline{Z}$ for the XAS measurements and provide important constraints in phase space for modeling these plasmas. 
  
   Importantly, with this modeling approach we can estimate the plasma conditions, subject to model-dependent uncertainties, and thereby place meaningful constraints on the treatment of density-dependent effects and outer-shell ionization in warm dense copper. In this regime, where the thermal, Coulomb, and Fermi energies are of comparable magnitude, a self-consistent description of bound states and the continuum remains a significant challenge for current warm dense matter models.

  In the measured spectra, we observe systematic blue shifts of the K-edge by 7 to 30 eV (from low to high drive) and shifts of 4 to 26 eV in the bound 1s$\rightarrow$3p resonance relative to the cold edge. Figure \ref{fig:csd_shifts}(b) shows the three experimental K-edge shifts as a function of their inferred $\overline{Z}$ (Nguyen). Taken together, these measured shifts, combined with the inferred temperature, density, and charge-state distribution (CSD), provide a compact and experimentally anchored data set for direct comparison with ionization and opacity models, and for benchmarking future theoretical and experimental studies in this regime. Table \ref{tab:summary} summarizes the inferred plasma conditions for the three drive configurations.
   
  From the measured XAS data, we identify the plasma conditions that best represent the experiment while explicitly accounting for the influence of model-dependent effects on the inferred parameters. Although both FAC screening approaches predict similar CSDs within experimental uncertainties, neither reproduces the measured edge energy, $E_{edge} = E_{1s} + \mu(n_{e}, T)$. This discrepancy, while not unexpected, highlights specific limitations in current treatments of continuum lowering and electronic structure at these densities and temperatures. With additional experimental constraints, it may be possible to decompose the K-edge into distinct contributions from bound-state energies, the chemical potential, and ionization potential depression (IPD). Achieving this would require validated calculations of the binding energies with FAC, together with an independent and accurate determination of the electron density, which remains an experimental challenge for this platform and for WDM research more broadly.
    
    \begin{table}
      \begin{ruledtabular}
      \begin{tabular}{|c|c|c|c|}
        Exp. inferred values & High-drive & Mid-drive	& Low-drive \\
        \hline
        T	[eV] & 20.8$\pm$0.7 & 15.7 $\pm$ 0.6 & 10.6 $\pm$ 0.4 \\
        $\overline{Z}$ (Nguyen) & 6.9 $\pm$ 0.8	& 5.9 $\pm$ 0.8 & 4.0 $\pm$ 0.8 \\
        K-edge shift [eV] & 29.4 $\pm$ 4.3 & 17.8 $\pm$ 4.4 & 6.7 $\pm$ 3.5 \\
        1s$\rightarrow$3p shift [eV] & 25.7 $\pm$ 4.4 & 12.6 $\pm$ 4.4 & 3.9 $\pm$ 3.5 \\
        \hline \hline
        from HYDRA & High-drive & Mid-drive	& Low-drive \\
        \hline
        $\rho$ [g/cm$^3$] & 18 $\pm$ 4 & 24 $\pm$ 3 & 24 $\pm$ 3 \\
      \end{tabular}
      \end{ruledtabular}
      \caption{\label{tab:summary} Summarized inferred conditions from the three drive configurations in this study. Density is inferred from experimentally verified simulations; temperature and ion charge are experimentally inferred; K-edge and K$\beta$ shifts are reported in reference to the cold ambient Cu K-edge.}
    \end{table}

\section{\label{sec:conclude}Conclusions}

    We have presented experimental results inferring ionization and temperature from high-quality x-ray absorption spectroscopy (XAS) measurements of uniform warm dense copper plasmas with minimal spatial and temporal gradients. By varying the laser drive intensity, we observe clear sensitivity of the K-edge region to changes in temperature and ionization. Across three different drive conditions, we infer temperatures of 10.6 $\pm$ 0.4 eV,  15.7 $\pm$ 0.6 eV, 20.8$\pm$0.7 eV and average ionization $\overline{Z}$ of 4.0 $\pm$ 0.8, 5.9 $\pm$ 0.8, and 6.9 $\pm$ 0.8, respectively, using the Nguyen PPS for the low-, mid-, and high-drive configurations.

    The temperature and ionization of the plasma are inferred using a functional fit to the K-edge and the bound-bound resonance absorption feature. Under these degenerate conditions, the K-edge is assumed to be primarily shaped by the Fermi-Dirac distribution of the electron population, allowing the temperature to be inferred directly from the edge slope. This temperature inference provides a critical constraint for studies of ionization in this regime. We further demonstrate that fitting individual charge-state contributions to the bound-bound resonance absorption feature enables inference of the plasma charge-state distribution. While the choice of atomic model leads to variations in the inferred values, these differences provide a useful bound for comparison with more sophisticated models that currently struggle to reproduce the measured spectra. Experimental data in this WDM regime are therefore essential for improving the understanding of density-dependent effects on the atomic structure of embedded ions and for constraining future modeling efforts.

    Owing to its relative simplicity, this experimental platform is readily adaptable to studies of other mid-Z materials and to the inclusion of complementary diagnostics. Future experiments could further constrain plasma conditions by incorporating additional spectroscopic measurements, such as x-ray fluorescence spectroscopy, which in conjunction with XAS can probe ionic structure through emission spectra. In the present measurements, source-size broadening limited the achievable spectral resolution, and improvements in this area may yield additional insight into line-broadening mechanisms. Continued experimental investigations of partially ionized metals in the warm dense regime will be essential for addressing persistent discrepancies in models of ionization potential depression and other dense plasma effects.

\section{\label{sec:acknowledge}Acknowledgements}
  This work was performed under the auspices of the U.S. Department of Energy by Lawrence Livermore National Laboratory under Contract DE-AC52-07NA27344 and was supported by the LLNL-LDRD Program under Project No. 22-ERD-005. SNL is managed and operated by NTESS under DOE NNSA contract DE-NA0003525. This work is partially supported by the Department of Energy, National Nuclear Security Administration under Award No. DE-NA0004147. This material is based upon work supported by the Department of Energy [National Nuclear Security Administration] University of Rochester “National Inertial Confinement Fusion Program” under Award Number(s) DE-NA0004144. SNL is managed and operated by NTESS under DOE NNSA contract DE-NA0003525.

\appendix

\section*{Appendix}
\section{\label{appendix:efx_res}Spectral Resolution of the EFX Spectrometer}
  Quantifying the spectral resolution of the EFX spectrometer is essential for the analysis and fitting of the copper K-edge data. Following the model outlined by Chin et al. \cite{chin_high-resolution_2023} the total spectral resolution is decomposed into four contributions: source size broadening, the crystal rocking-curve contribution, detector resolution, and depth broadening due to x-ray penetration into the crystal. Each term is represented as the FWHM of a Gaussian:

\begin{align} 
    \delta E_{\rm{src}} &= \frac{dE}{dx}\frac{D_{\rm{src}}}{\rm{cos}(\pi/2 - \theta_{B})}\\
    \delta E_{\rm{rock}} &= E\frac{\delta_{rock}(E)}{\rm{tan}(\theta_{B})}\\
    \delta E_{\rm{det}} &= \frac{dE}{dx}\chi_{\rm{ip}}\\
    \delta E_{\rm{depth}} &= \frac{dE}{dx}2l_{\rm{depth}}(E)\rm{cos}(\theta_{\rm{B}})
\end{align}

  Here, $dE/dx$ is the change in unit energy per distance along the detector derived from the dispersion relation. $D_{\rm{src}}$ is the backlighter source size (140 $\mu$m for these experiments), $\delta_{rock}$ is the FWHM of the calculated rocking curve as a function of energy, $\chi_{\rm{ip}}= 25$ $\mu$m is the image plate resolution, and $l_{\rm{depth}}$ is the energy- and angle-dependent x-ray penetration depth in silicon calculated using NIST data\cite{henke_x-ray_1993}. The total spectral resolution is obtained by adding the four contributions in quadrature:

\begin{equation}
  \delta E_{\rm{total}}=\sqrt{\delta E_{\rm{src}}^2 + \delta E_{\rm{rock}}^2 + E_{\rm{det}}^2 + E_{\rm{depth}}^2}
\end{equation}

\begin{figure}
  \centering
  \includegraphics[width=1.0\linewidth]{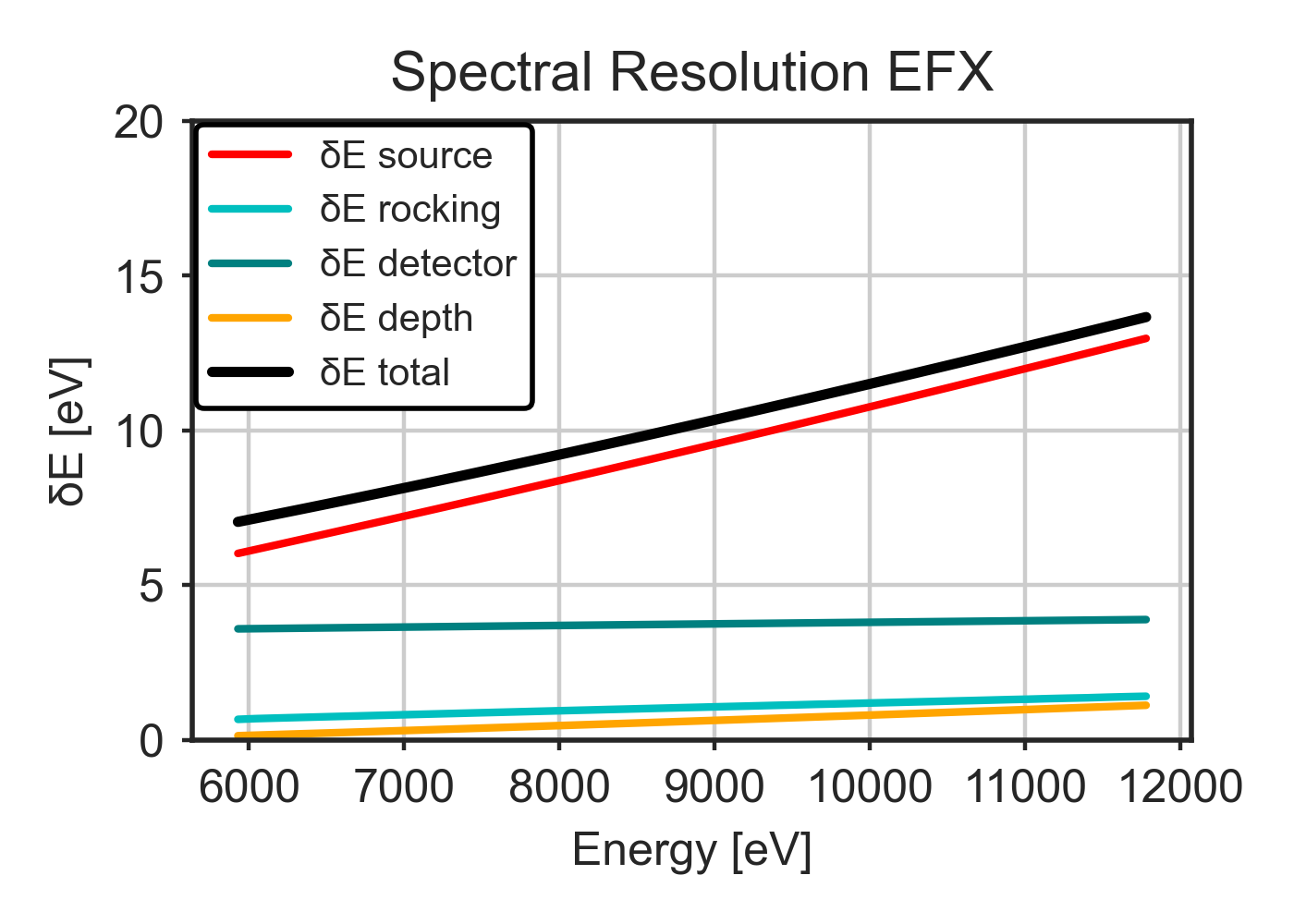}
  \caption{The spectral resolution for each term along with the total resolution in black. The largest contribution comes from the source size, \qty{140}{\micro\meter} for our experiments. Additional contributions come from broadening due to the rocking curve, the image plate resolution, and the x-ray penetration depth.}
  \label{fig:efx_res}
\end{figure}

  The individual contributions to the spectral resolution, as well as the total calculated resolution, are shown in Figure \ref{fig:efx_res} for the EFX configured for the Cu XAS experiments at OMEGA. Source size broadening is the dominant contributor to total resolution. In the region of interest around the Cu K-edge, we find a resolution of approximately 10 eV. This is in agreement with our findings from the undriven cold Cu K-edge measurements (see Fig \ref{fig:ambanddrive}(a)).

\begin{figure}
  \centering
  \includegraphics[width=1.0\linewidth]{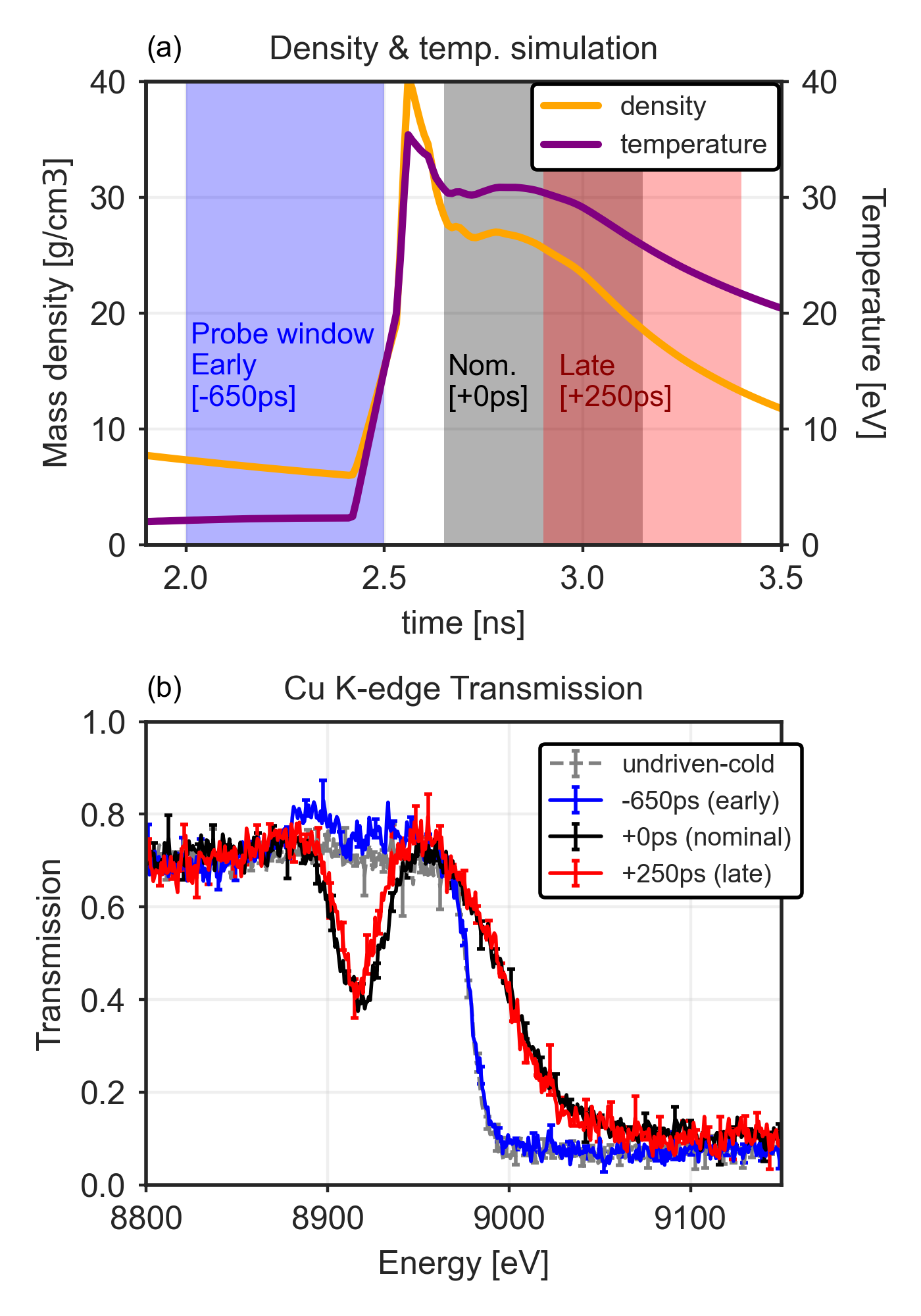}
  \caption{(a) HYDRA simulation of density and temperature in the buried Cu layer overlaid with backlighter probe times to measure XAS early timing (blue), nominal timing (black), and late timing (red). (b) The resulting experimental XAS data for the three timing variations are compared to the undriven cold case (gray). We find little variation in the K-edge, in the early timed case, confirming the timing and similar results in the nominal and late timed, confirming the nominal is measuring after the shock compression only.}
  \label{fig:cuxfs_timing}
\end{figure}

\section{\label{appendix:cuxfs_timing}Experimental Timing Scan of conditions}

  Additional experiments were performed to verify the shock timing inferred from the one-sided VISAR measurements. In these experiments, the platform was configured in the nominal two-sided geometry (see Fig. \ref{fig:omegaplatform}). The x-ray backlighter timing was varied to measure XAS at three characteristic times: prior to shock arrival at the buried copper layer, at the nominal optimal time shortly after complete shock propagation through the copper, and at a delayed time relative to this optimum. This timing scan provides an experimental validation of the HYDRA-simulated timing used to design the primary experiments.

  Figure \ref{fig:cuxfs_timing} shows representative HYDRA simulations and corresponding XAS measurements from three shots with varied backlighter timing for the two-sided mid-drive configuration (2 ns laser drive). The early-time shot (blue) probes a 500 ps window before the shocks reach the buried copper layer. The nominal shot (black) is timed to the optimal window just after peak compression of the buried layer, while the late-time shot (red) is delayed by an additional 250 ps relative to the nominal timing. Figure \ref{fig:cuxfs_timing}(a) compares the three backlighter timing windows to the HYDRA-simulated density evolution within the buried copper layer, and Fig. \ref{fig:cuxfs_timing}(b) shows the corresponding measured XAS in the Cu K-edge region.

  Based on the HYDRA simulations, the early-time shot is expected to probe near-ambient copper conditions, which is confirmed experimentally by the observed K-edge being identical to that of the undriven case. For both the nominal and late-time shots, the simulations predict probing only warm dense copper. The close similarity of the measured XAS spectra for these two shots indicates that the nominal timing window occurs after complete shock propagation through the copper layer, as the presence of any unshocked material would produce readily identifiable changes in the absorption spectrum. In addition, the early-time shot indicates that laser-driven preheat may produce detectable K$\beta$ fluorescence, but does not significantly heat the buried copper layer, as no K-edge softening associated with thermal effects is observed.

  Together, these measurements demonstrate that the HYDRA simulations, once adjusted to match the one-sided VISAR data, provide sufficient accuracy for timing the x-ray backlighter to probe the optimal window just after full shock traversal of the buried copper layer.

  \bibliography{CuXAS_Bibliography.bib}

\end{document}